\input epsf
\epsfverbosetrue

\documentstyle[LTpaper]{article}
\begin{document}
\title{Resistive transport in a mesoscopic proximity superconductor} 
\author{P. Charlat$^a$, H. Courtois$^a$, Ph. Gandit$^a$, D. Mailly$^b$, 
A.F.~Volkov$^c$, and B. Pannetier$^a$}
\address{\mbox{$^a$ C.R.T.B.T.-C.N.R.S., Laboratoire associ\'e \`a l'universit\'e 
Joseph Fourier,}\\
\mbox{25 Av. des Martyrs, 38042 Grenoble, France}
\vskip 10pt
\mbox{$^b$ L.M.M.-C.N.R.S.,}\\
\mbox{196 Av. H. Ravera, 92220 Bagneux, France}
\vskip 10pt
\mbox{$^c$ Institute of Radio Engineering and Electronics,}\\
\mbox{Russian Academy of Sciences, Mokhovaya st.11, 103907 Moscow, Russia}}

\abstract{We review transport measurements in a normal metal (N) in contact with one 
or two superconducting (S) islands. From the experiment, we distinguish the 
Josephson coupling, the mesoscopic fluctuations and the proximity effect. In a 
loop-shaped N conductor, we observe large $h/2e$-periodic magnetoresistance 
oscillations that decay with temperature $T$ with a $1/T$ power-law. This behaviour 
is the signature of the long-range coherence of the low-energy electron pairs induced 
by the Andreev reflection at the S interface. At temperature and voltage below the 
Thouless energy $\hbar D / L^2$, we observe the re-entrance of the metallic 
resistance. Experimental results agree with the linearized quasiclassical theory.}
\maketitle
\pagestyle{empty}

\section{Introduction}
Proximity effect between a normal metal (N) and a superconductor (S) is a unique tool 
for studying the interplay between dissipation and quantum coherence on one hand, 
resistive transport 
and superconductivity on the other hand. Induced superconductivity in a normal metal 
manifests itself by the enhancement of the local conductivity compared to the intrinsic 
metallic regime, the Josephson effect in a S-N-S junction and the appearance of a 
(pseudo-) gap in the density of states. Although all these effects have been observed 
and understood in the early 60's, recent experimental work triggered a large interest in 
S-N structures \cite{Karlsruhe}. This revival is mostly due to the mesoscopic size and 
the well-controlled design of the recently investigated samples.

On the microscopic scale, superconductivity diffuses in the normal metal thanks to 
the Andreev reflection of electrons at the S-N interface \cite{Andreev}. In this 
process, an electron coming from N is reflected as a hole of opposite momentum, 
while a Cooper pair is transmitted in S. In consequence, this process correlates 
electron states of the same energy $\epsilon$ compared to the Fermi level $E_F$, but 
with a slight change in wave-vector $\delta k$ due to the branch crossing : $\delta k / k_F 
= \epsilon/E_F$, $k_F$ being the Fermi wavevector. After elastic diffusion up to a 
distance L from the interface, the wave-vector mismatch induces a phase shift $\delta 
\varphi = (L/L_{\epsilon})^2  = 
\epsilon/\epsilon_{c}$. In consequence, a pair maintains coherence in N up to the
 energy-dependent length $L_\epsilon = \sqrt{\hbar D/\epsilon}$. Taking 
$2 \pi k_B T$ as a characteristic energy of the incident electron, one finds again the 
well-known thermal length $L_T = \sqrt{\hbar D/2 \pi k_B T}$ in the "dirty" limit $l_p < 
L_T$. This length is indeed described in the textbooks \cite{deGennes} as the 
characteristic decay length of the proximity effect.

If one measures the resistance of a 
long S-N junction, the length $L_T$ appears missing, so to say superconducting. Does 
it mean that superconductivity in N is restricted to a region of length $L_T$ from the 
interface ? The answer is clearly no. In fact, it is easy to see that at the Fermi level 
($\epsilon = 0$), the value of the diffusion length $L_\epsilon$ diverges. The coherence 
of low-energy electron pairs is then only bound by the phase-coherence of one electron, 
which persists on the length $L_\varphi$. In a pure metal at low temperature, the 
phase-coherence length $L_{\varphi}$ can reach several microns 
\cite{Pannetier-Rammal} so that it is much larger than the thermal length $L_T$. In this 
scope, the thermal length $L_T$ and the phase-coherence length $L_{\varphi}$ are the 
smallest and the largest of the coherence lengths $L_\epsilon$ ($\epsilon < k_B T$) of the 
electron pairs present at the temperature $T$. The strong coherence at the Fermi level 
justifies the survival up to $L_{\varphi}$ of a local proximity effect on the resistive 
transport that can be tuned by an Aharonov-Bohm flux 
\cite{Petrashov,Dimoulas,CourtoisPrl}.

In the quasiclassical theory \cite{Volkov,Zaitsev,Zhou}, the phase conjugation 
between the electron and the hole is expressed as a finite pair amplitude involving states 
of wave-vectors $(k_F+ \epsilon / \hbar v_F,-k_F+ \epsilon / \hbar v_F)$, $v_F$ being 
the Fermi velocity. Looking back at the expression of the wave-vector mismatch $\delta 
k$, one can see that at a distance $L$ from the interface, electron-hole coherence is 
restricted to an energy window of width the Thouless energy $\epsilon_{c}=\hbar 
D/L^2$. The pair amplitude $F(\epsilon,x)$ as a function of both energy and distance 
from the S interface therefore varies on the scales of $\hbar D/x^2$ and $L_\epsilon$ 
respectively. As a result, the latter energy enters as a characteristic energy in the local 
density of states of the N conductor \cite{Gueron} at large distance $x > \xi_s$. The 
presence of a finite pair amplitude $F(\epsilon)$ also induces a local and 
energy-dependent enhancement of the effective conductivity of N. A surprising feature 
appears at low temperature, when $L_T$ is larger than the sample size $L$, and zero 
voltage. In this regime, electron pairs are coherent over the whole sample, but diffusion 
of individual electrons is predicted to be insensitive to the presence of electron pairs 
\cite{Stoof,Volkov-Lambert}.

In this paper, we review the different contributions to the phase-coherent transport in a 
mesoscopic N metal with S electrodes. We show under which conditions the 
Josephson coupling, conductance fluctuations, and proximity effect prevail. Various 
sample geometry's have been used in order to precisely measure the different 
quantities. We will demonstrate that the relevant energy for the Josephson coupling in 
a {\it mesoscopic} S-N-S junction is not the superconducting gap $\Delta$, but the Thouless energy $\epsilon_c$. In the resistive state, we find a phase-sensitive contribution to transport carrying a $1/T$ power-law decay with 
temperature. This proximity effect is much larger than the weak localization 
correction \cite{Spivak} and results from the persistence of electron-hole coherence far 
away from the S-N interface. Eventually, we report new experimental results showing for 
the first time the low-temperature re-entrance of the metallic resistance in a mesoscopic 
proximity superconductor. This re-entrant regime is destroyed by increasing the 
temperature \cite{Stoof} or the voltage \cite{Volkov-Lambert}. An Aharonov-Bohm flux 
can modify the sample effective length and therefore shift the energy crossover. All these results are shown to be in qualitative agreement with the quasiclassical theory.

\section{A theoretical approach to S-N junctions}
Superconductor--Normal-metal structures can be well described by the 
quasiclassical theory for inhomogenous superconductors. We present here a simplified 
version that however keeps the essential physical features. Let us consider the 
mesoscopic regime where the inelastic scattering length is larger than the sample length. 
The flow of electrons at some energy $\epsilon$ is then uniform over the sample, so that 
one has to consider parallel and independent channels 
transporting electrons at a particular energy $\epsilon$. In a perfect reservoir, electrons 
follow the Fermi equilibrium distribution at the temperature $T$ and chemical potential 
$\mu$. Charges are injected in the system from one reservoir at $\mu = eV$ and 
transferred to the other, so that the current is carried by electrons within an energy 
window $[0,\mu = eV]$ with a thermal broadening $k_B T$.

In N, $F(\epsilon,x)$ follows the Usadel equation that for small $F$ can be linearized as 
: 
\begin{equation} 
\hbar D \partial^{2}_{x} F + (2 i\epsilon - \frac{\hbar D}{L_{\varphi}^2}) F = 0
\label{diffusion}
\end{equation}
The linearization used above is a very crude approximation, since near the interface 
which is believed to be clean, the pair amplitude should be large (in this 
approximation, we have $F(\epsilon \ll \Delta,0) =-i\pi /2$ for a perfectly transparent 
interface \cite{Charlat}). At the contact with a N reservoir $F$ is assumed to be zero. 
However, this simple formulation enables a straightforward understanding of the 
physical root of the proximity effect in a S-N junction. Eq. 1 features simply a 
diffusion equation for the pair amplitude $F$ at the energy $\epsilon$ with a decay 
length $L_\epsilon$ and a cut-off at $L_\varphi$. At a particular energy $\epsilon$, 
the imaginary part is maximum at the S-N interface and decays in an oscillating way 
on the length scale of $L_\epsilon$ if $L_\epsilon \ll L$. The real part of the pair 
amplitude is zero at the S interface, maximum at a distance $L_\epsilon$ if $L_\epsilon \ll 
L$ and then decays also in an oscillating way. 

From the theory, one can calculate everything, including the Josephson effect in a 
S-N-S geometry \cite{Zaikin}, the change in the density of states \cite{Gueron} and 
the conductivity enhancement \cite{Volkov}. Here we will concentrate on the last point. 
The pair amplitude $F$ is responsible for a local enhancement of the conductivity 
$\delta \sigma(\epsilon,x) = \sigma_N (Re [F(\epsilon,x)])^{2}$ for small $F$, 
$\sigma_N$ being the normal-state conductivity \cite{Volkov,Zhou,Stoof}. Part of this 
contribution is similar to the Maki-Thompson contribution to the fluctuation in a 
conventional superconductor \cite{MT}. The other part is related to a modification of the 
density of states and compensates the first one at zero energy \cite{Volkov}.

From the behaviour of $\delta \sigma(\epsilon,x)$ it is then 
straightforward to calculate the excess conductance $\delta g(\epsilon )$ for the precise 
geometry of the sample. The measured excess conductance $\delta G(V,T)$ at voltage 
V and temperature T writes :
\begin {equation}
\delta G(V,T) = \int_{-\infty}^{\infty}\delta g(\epsilon ) P(V-\epsilon)d \epsilon 
\label {total}
\end {equation}
where $P(\epsilon)=[4k_B T ${cosh}$^2(\epsilon/2k_B T)]^{-1}$ is a thermal kernel 
which reduces to the Dirac function at $T = 0$. Hence, the low-temperature 
differential conductance $dI/dV = G_{N}+\delta G$ probes the proximity-induced excess 
conductance $\delta g(\epsilon)$ at energy $\epsilon = eV$ with a thermal broadening 
$k_{B}T$.

As an example, let us consider a long N wire connected to a S reservoir at one end. Along 
the wire, the excess conductivity $\delta \sigma(\epsilon,x)$ at a given energy $\epsilon $ 
increases from zero at the S-N interface to a maximum of about 0.3 $\sigma_N $ at a 
distance $L_\epsilon$ from the interface (if $L \gg L_{\epsilon}$) and then decays 
exponentially with $x$. The integrated excess conductance $\delta g(\epsilon)$ of the 
whole sample rises from zero with an ${\epsilon}^2$ law at low energy, reaches a 
maximum of 0.15 $G_N$ at about 5 $\epsilon_c$ and goes back to zero at higher energy 
with a $1/\sqrt{\epsilon}$ law. In the regime $L > L_T$, this contribution is responsible 
for the subtraction of a length $L_T$ in the resistance of the sample. The effect is 
maximum when temperature or voltage is near the Thouless energy : $k_B T$ or $eV 
\simeq \epsilon_c$, i.e. when $L \simeq L_T$. At zero voltage and temperature ($eV, 
k_B T \ll \epsilon_c$), the conductance enhancement is predicted to be zero, since $Re 
[F(0,x)] = 0$ everywhere.

\section{Sample fabrication and characterization}
We fabricated S-N samples with a clean S-N interface and various geometry's for the 
N conductor. Two different techniques were used, depending on the desired geometry. In 
the case of long N wires, oblique metallic evaporation's were performed in a 
Ultra-High Vacuum (UHV) chamber through a PMMA resist mask suspended 500 nm 
above a Si substrate. During the same vacuum cycle, N = Cu or Ag 
and S = Al structures are evaporated with the same angle of 45$^\circ$ but along perpendicular axes of the substrate plane \cite{Courtois}. The whole process lasts about 5 
minutes with a pressure never exceeding 2.$10^{-8}$ mbar, which ensures us of a 
high-quality interface between the two metals.

A different technique has been used for nanofabricating single Cu loops with or two Al 
electrodes. The Cu and the Al structures are patterned by conventional lift-off e-beam 
lithography in two successive steps with repositioning accuracy better than 
100 nm. In-situ cleaning of the Cu surface by 500 eV $Ar^+$ ions prior to Al 
deposition ensures us of a transparent and reproducible interface. In this case, the base 
pressure is about $10^{-6}$ mbar. The typical interface resistance of a S-N junction 
of area 0.01 $\mu m^2$ was measured to be less than 1 $\Omega$.

We performed transport measurements of a variety of samples in a $\mu$-metal--shielded 
dilution refrigerator down to 20 mK. Miniature low temperature high-frequency filters 
were integrated in the sample holder \cite{Filtres}. Sample 1 is made of a continuous Ag 
wire in contact with a series of Al islands of period 800 nm, see Fig. 1. The 
width and thickness of the Ag wire are 210 nm and 150 nm respectively, of the Al wire 
120 nm and 100 nm. The normal-state resistance of one junction $r_n =$ 0.66 $\Omega$ 
gives the elastic mean free path in Ag : $l_p =$ 33 nm, the diffusion coefficient $D = v_F 
l_p /3$ = 153 $cm^2/s$ and the thermal length $L_T =$ 136 nm at 1 K.

Due to the strong proximity effect in Ag below each Al island, the whole sample can be 
considered as a 1D array of about one hundred S-N-S junctions. Below the 
superconducting transition of Al at $T_{c} \simeq 1.4$ K, the resistance of sample 1 
indeed decreases by an amount $(10 \%)$ compatible with the coverage ratio of the Ag 
wire by Al islands $(\simeq 15 \%)$. This behaviour is in agreement with a low S-N 
interface resistance and shows that superconductivity in Al is little depressed by the 
contact with Ag. Let us point out that we do not observe any resistance anomaly or peak 
at $T_c$ as soon as the current-feed contacts are in-line with the sample \cite{Moriond}. 
At low temperature ($T <$ 800 mK), the resistance drops to zero due to the Josephson 
effect. The appearance of a pair current here, in contrast with previous experiments on 
similar samples \cite{Petrashov}, should be attributed to a much cleaner interface thanks 
to the UHV environment.

\section{Josephson coupling}
In all our samples, the length $L$ of the N conductor between two neighbouring S 
islands is larger than the superconducting coherence length $\xi_s$ and smaller than 
the phase-breaking length $L_\varphi$ : $\xi_s < L < L_\varphi$. In terms of energy, the 
first relation also means that the Thouless energy $\epsilon_c$ is smaller than the 
superconducting gap $\Delta$. We call this regime mesoscopic, as opposed to the 
classical regime $L < \xi_s$, in which the two S electrodes are strongly coupled. In this 
regime, one could expect (since $L > \xi_s$) a depletion of superconducting properties 
but (since $L < L_\varphi$) still a strong quantum coherence.

\vspace*{0.5 cm}
\epsfxsize=8 cm 
\begin{center}
\small 
{\bf Fig. 1} : Temperature dependence of the critical current of sample 1 in 
parallel with the fit derived from the de Gennes theory in the dirty limit $l_p < L_T = 
\sqrt{\hbar D /k_B T}$, which is relevant for the sample. Inset : oblique view of a similar 
sample made of a continuous Ag wire of width 210 nm in contact with a series of Al 
islands. The distance between two neighbouring Al islands is 800 nm.
\normalsize
\end{center}

We studied extensively the behaviour of the critical current as a function of 
temperature in a variety of samples \cite{Courtois}. Fig. 1 shows 
the critical current of sample 1, which is representative of many other samples 
consisting of one or many S-N-S junctions in series. The first remarkable point is that the 
low temperature saturation value of the $r_N I_c$ product of about 1.7 $\mu V$ is much 
smaller than the energy gap ($\Delta \simeq$ 300 $\mu V$) of Al. This is in clear 
contrast with what is observed in S-I-S tunnel junctions and short S-N-S junctions 
\cite{Jct}. Moreover, the experimental data as a function of temperature does not fit 
with the classical de Gennes theory \cite{deGennes} derived from the Ginzburg-Landau 
equations, see Fig. 1. In order to get a good fit, one has to take the clean limit ($l_p > 
L_T = \hbar v_F / k_B T$) expression for $L_T$ \cite{Courtois}, but even in this 
non-relevant regime the fit parameters ($v_F$ and $\Delta$) are not correct.

These results motivated a recent theoretical study \cite{Zaikin} that met most of the 
experimental facts. The dirty limit is considered. In the mesoscopic regime where $L > 
\xi_s$, it is found that the relevant energy is no 
longer the gap $\Delta$ but the Thouless energy $\epsilon_c$. This is indeed consistent 
with our physical argument that only electron pairs within the Thouless window around 
the Fermi level remain coherent over the N metal length $L$. In this case, the $r_N I_c$ 
product at zero temperature is equal to $\epsilon_c / 4$. This result was already 
known in the ballistic regime \cite{Bagwell}. Eventually, the 
calculated temperature dependence differs strongly from the naively expected 
$exp(-L/L_T)$ law as soon as all Matsubara frequencies are taken into account. The 
argument below the temperature in the exponential is indeed proportional to the Thouless 
energy \cite{Zaikin}. A good agreement of the theory with Fig. 1 experimental points has 
been obtained by taking $L =$ 1.2 $\mu m$ \cite{Zaikin}. The discrepancy with the 
actual size of 0.8 $\mu m$ may at least partly be attributed to the crude modelization of 
the sample geometry.

Sample 2, see Fig. 3 inset, is made of a single Cu loop in contact with two S islands. 
Diameter of the loop is 500 nm, width and thickness of the Cu wire are 50 nm and 25 
nm. Centre-to-centre distance between the 150 nm wide Al islands is 1 $\mu m$. The 
normal-state resistance $R_N =$ 51 $\Omega$ gives a mean free path $l_{p} 
=$ 16 nm, a diffusion constant $D =$ 81 $cm^2/s$ and a thermal length $L_{T}=$ 
99 nm/$\sqrt{T}$. The amplitude of the conductance fluctuations measured in the purely 
normal state (not shown) gives the value of the phase-coherence length : $L_{\varphi} =$ 
1.9 $\mu m$, so that the whole structure is coherent.

Near $T_c$, sample 2 resistance behaves very closely to 
that of sample 1. At low temperature ($T < 250$ mK), sample 2 resistance drops to 
a constant value of 16 $\Omega$ due to the Josephson coupling between the two S 
electrodes. The residual resistance can be related to the resistance of the normal metal 
between each voltage contact and the neighbouring S island, with an extra contribution 
due to the current conversion. In this truly mesoscopic sample, the critical current 
behaves very like in the long N wires with a series of S islands. In addition, the critical 
current is modulated by the magnetic flux $\phi$ with a period $\phi_{0}$ 
\cite{CourtoisPrl}. This behaviour is reminiscent of a superconducting quantum 
interference device (SQUID), although our mesoscopic geometry differs strongly from 
the classical design. 

\section{The conductivity enhancement}
In the temperature regime $L_T \ll L$, which corresponds in sample 2 to $T >$ 500 
mK, the pair current is exponentially small but still sensitive to the magnetic flux. 
Thermal fluctuations should therefore induce exponentially-small magnetoresistance 
oscillations. From the classical RSJ model \cite{RSJ} one can extrapolate a relative 
amplitude of $10^{-8}$ at 1 K. In contrast, the experiment (see Fig. 2) exhibits 
pronounced magnetoresistance oscillations with a relative amplitude of 0.8 $\%$ at 1 K. 
The only flux-periodicity is ${\it h}/2e$ as no structure of half periodicity 
was met, even when the measuring current amplitude was changed. We used small 
currents, so that electrons chemical potential difference $eV$ between the two N 
reservoirs was less than the Thouless energy. In this "adiabatic" regime, the phase 
difference between the two S electrodes can be considered as constant despite the finite 
voltage drop $V$ between them.

\vspace*{0.5 cm}
\epsfxsize=8 cm \epsfbox{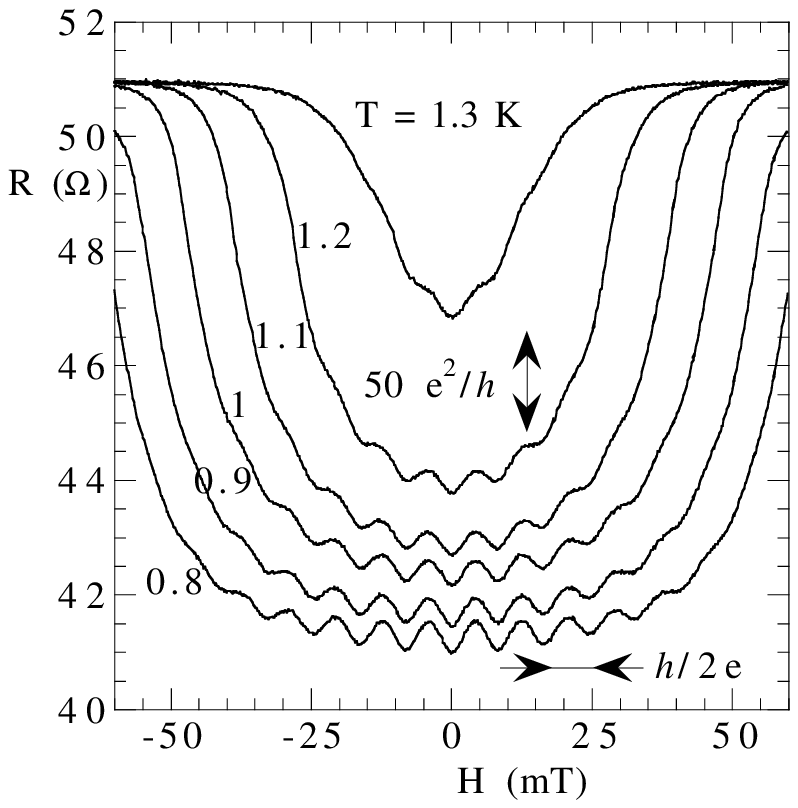}
\begin{center}
\small 
{\bf Fig. 2} : Sample 2 low-field magnetoresistance for $T =$ 0.7; 0.8; 0.9; 1; 1.1; 1.2 
and 1.3 K with a measurement current $I_{mes} =$ 60 nA. $T =$ 0.7 and 0.8 K curves 
have been shifted down by 1 and 0.5 $\Omega$ respectively for clarity. Oscillations of 
flux-periodicity ${\it h}/2e$ and relative amplitude 0.8 $\%$ or equivalently 4.2 $e^2/{\it h}$ at $T =$ 1 K are visible.
\normalsize
\end{center}

The observed magnetoresistance oscillations are very robust, since they remain clearly 
visible very near $T_c$ at 1.3 K, when the thermal length $L_T$ is much smaller than 
the distance between one Al island and the loop. Fig. 3 shows the temperature 
dependence of the two main contributions to transport. The Josephson current vanishes 
rapidly above 250 mK, revealing the exponential decay with temperature. The amplitude 
of the observed h/2e magnetoresistance oscillations is plotted on the same graph. It shows 
a striking agreement with a fit using a plain $1/T$ power-law. The slight deviation of the 
data from the $1/T$ fit near $T_{c}$ is clearly related to the depletion of 
superconductivity in S. This power-law dependence is a new result, in clear contrast with 
the exponential damping with the temperature of the Josephson current. The observation 
of the same effect in samples with only one S island, see below, further illustrates the 
difference with the Josephson effect. Eventually, the large amplitude of the effect 
compared to the quantum of conductance $e^2/{\it h}$ discards an interpretation in terms 
of weak localization or conductance fluctuations.

\vspace*{0.5 cm}
\epsfxsize=8 cm \epsfbox{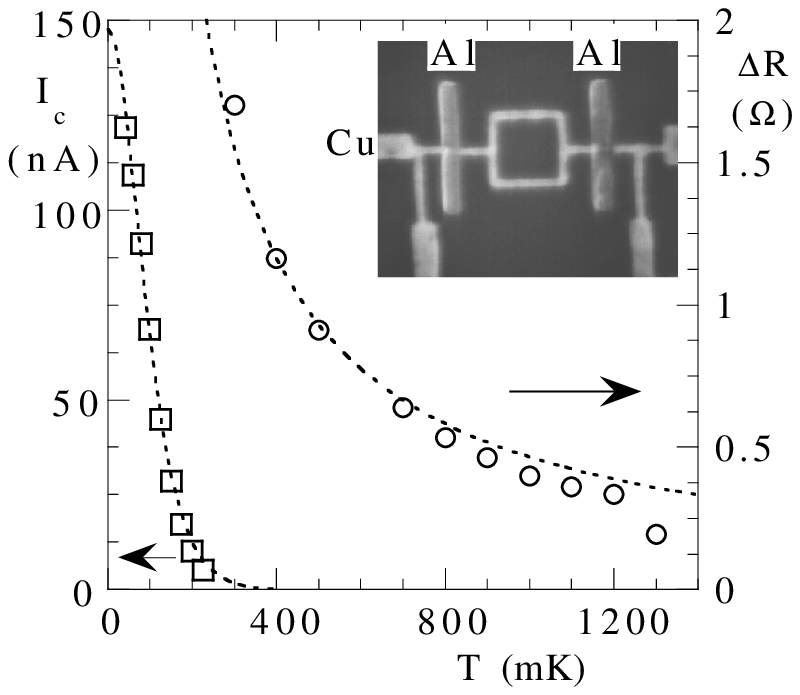}
\begin{center}
\small 
{\bf Fig. 3}, Left scale : Temperature dependence of the critical current of sample 2 with a 
25 $\Omega$ differential resistance criteria. The dashed line is a guide to the eye. 
Right scale : Temperature dependence of the amplitude of the 
magnetoresistance oscillations, $I_{mes} =$ 60 nA. The dashed line is a $1/T$ fit. 
Inset : micrograph of a similar sample made of a single Cu in contact with two (here 
vertical) Al islands. The whole sample, including the measurements contacts and except 
the two Al islands, is made of Cu. The loop diameter is 500 nm.
\normalsize
\end{center}

In our geometry, the loop is a unique tool for selecting the long-range component of the 
proximity effect by making interfere the pair amplitudes of both arms of the loop. The 
observation of a power-law temperature dependence instead of the naively expected 
exponential damping over $L_T$ definitely shows that the natural length scale of the 
proximity effect is not the thermal length $L_T$ but the phase-memory length 
$L_\varphi$. The role of temperature is only to distribute the energy of the 
electrons that will reflect on the superconducting interface and contribute to the proximity 
effect. At the Fermi level, electrons in a pair are perfectly matched, so that only 
phase-breaking can break the pair. It is quite remarkable that the relative 
amplitude of the oscillations is in agreement with the ratio $\epsilon_c / k_B T =$ 1.5 
$\%$ at 1 K if one takes the characteristic length $L =$ 2 $\mu m$ for the sample. This 
ratio can be seen as the fraction of electron pairs remaining coherent after diffusion up to a 
distance $L$ from the S-N interface.

\vspace*{0.5 cm}
\epsfxsize=8 cm \epsfbox{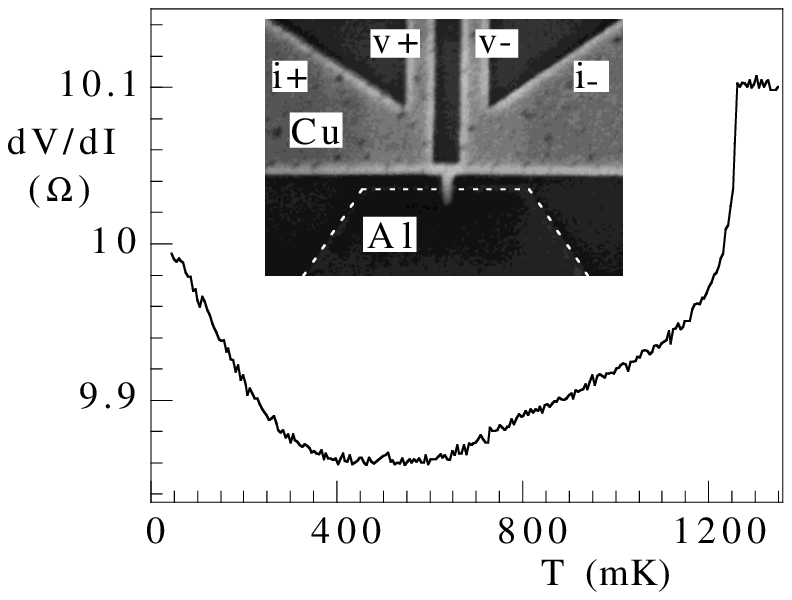}
\begin{center}
\small 
{\bf Fig. 4}: Temperature dependence of sample 3 resistance. Sample 3 is made 
of a single short N wire in lateral contact with a S island. Critical temperature of S = Al is 
1.25 K. Measurement current is 400 nA. Inset : micrograph of a similar sample made of a 
400-nm--long Cu wire between two massive N = Cu reservoirs and in lateral contact with 
a massive Al island. Due to the lack of contrast, a dotted line was drawn following the 
contour of the Al reservoir.
\normalsize
\end{center}

\section{Re-entrance regime}
With two S islands, one cannot investigate the dissipative transport in the strong 
coherence regime $L_{T} \simeq L$, since the Josephson effect shorts the low-current 
resistance between the two S islands. Thus we studied transport in samples with 
only one S island. Sample 3 consists in a 400 nm-long N wire in lateral contact with a 
single S island located 200 nm away from the N conductor centre, see Fig. 4 
inset. The width and thickness of the Cu wire are 80 and 50 nm respectively. The 
normal-state resistance is $R_N =$ 10.1 $\Omega$, the mean free path is $l_p =$ 6 nm, 
the diffusion coefficient $D$ is 30 $cm^2/s$ and the thermal length $L_T$ is 60 nm at 1 
K. The Thouless energy is then equal to 12 $\mu eV$ and the Thouless temperature is 
$\epsilon_c / k_B =$ 142 mK if one takes a characteristic length $L =$ 400 nm for the 
sample.

\vspace*{0.5 cm}
\epsfxsize=8 cm \epsfbox{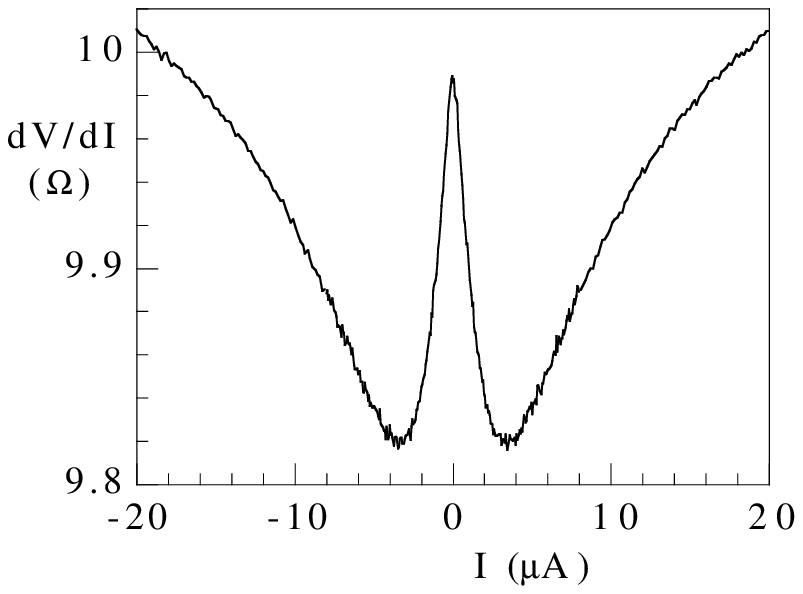}
\begin{center}
\small 
{\bf Fig. 5}: Bias current dependence of the differential resistance of sample 3 at $T =$ 50 
mK. Measurement current is 400 nA. The differential-resistance minimum corresponds to 
an electron energy of about 34 $\mu V$.
\normalsize
\end{center}

At low temperature, we do not observe a sharp drop of the resistance, since obviously no 
Josephson current can establish itself. In contrast, we observe an {\it increase} of the 
zero--magnetic-field and low-current resistance below 500 mK, see Fig. 4. 
The current bias dependence of the measured resistance shows the most striking 
behaviour with a {\it decrease} of the resistance when the polarisation current is 
increased, see Fig. 5. This non-linear behaviour discards 
an interpretation in terms of weak localization, which is known to be insensitive to 
voltage. The location of the resistance minimum gives a threshold voltage of 
about 34 $\mu V$. This value is of the order of the calculated value based on the 
Thouless energy $5 \epsilon_c =$ 61 $\mu V$. Moreover, the 500 mK crossover 
temperature for the temperature dependence is also compatible with the related Thouless 
temperature $5 \epsilon_{c} / k_B =$ 710 mK. In summary, we observed for the first 
time the re-entrance of the metallic resistance in a mesoscopic proximity superconductor. 
In agreement with the predictions \cite{Stoof,Volkov-Lambert}, this new regime occurs when all the energies involved are of order and below the Thouless energy of the sample.

Eventually, we studied electron transport in sample 4 that has exactly the same loop 
geometry than sample 2, except for the absence of one S island. The width of the 
wire is 150 nm, its thickness is about 40 nm, and the distance between the loop and 
the S island is about 100 nm. The normal state resistance $R_{N}$ = 10.6 $\Omega$ 
gives a mean free path $l_p =$ 18 nm, a diffusion coefficient $D = $ 70 $cm^2/s$ and a 
thermal length $L_T =$ 92 nm at 1 K. The behaviour of the resistance above 
500 mK is very similar to the two-island case : (i) At $T_{c}$ we observe a drop of the 
resistance of about 3 $\%$; (ii) Below $T_{c}$, the resistance oscillates with the 
magnetic field with an amplitude that follows again a $1/T$ temperature dependence 
down to 200 mK.

Fig. 6 (a) shows the temperature dependence of the resistance for various values of the 
magnetic flux in the loop. At zero-field, the re-entrance of the resistance is real but 
hardly distinguishable. At $\phi = \phi_0 / 2$, the re-entrance occurs at higher 
temperature $(T <$ 500 mK) and has a much larger amplitude. At $\phi = \phi_0$, the 
curve is close to the zero-field case, and at $\phi = 3 \phi_0 /2$, close to the  $\phi_0 / 
2$ case. With increasing further the magnetic field (Fig. 6 (a)), the re-entrant 
regime is entered at an increasing temperature, independently of the $\phi_0$-periodic 
modulation.

\section{Comparison with linearized theory}
Let us compare experimental results from sample 4 with calculation derived from the 
theory. Sample 4 can be modelized as two independent S-N circuits in series as 
shown in the inset of Fig. 6. This strong approximation describes the main 
physics of our particular geometry and illustrates more general situations. Both circuits 
consist of a N-wire between a superconductor S and a normal reservoir N. Because of the 
loop geometry, a magnetic field induces an Aharonov-Bohm flux, which changes the 
boundary conditions on the pair amplitude $F(\epsilon )$. At zero magnetic flux, 
$F(\epsilon )$ is zero only at the contact with the normal reservoirs. At half magnetic 
flux, destructive interference of the pair functions in the two branches enforces a zero in 
$F(\epsilon )$ at the node K (see Fig. 6 inset). Consequently, 
the pair amplitude is also zero between the loop and the N reservoir \cite{Charlat}. 
Half a flux-quantum then reduces the effective sample size to the length $L'$ between the 
S interface and the point K. As a result, the crossover temperature of the re-entrance of 
the resistance is shifted to higher temperature, see Fig. 6. Compared to the zero-field 
case, this subtracts the contribution of the pairs remaining coherent 
after the end of the loop, i.e. with an energy below the Thouless energy $\epsilon_{c}' 
= \hbar D / L'^2$. In the intermediate temperature regime $k_{B}T > \epsilon_{c}'$, 
this quantity is of the order of $\epsilon_{c}'/k_{B}T$. This is in qualitative agreement 
with the amplitude of the magnetoresistance oscillations in respect of the amplitude and 
the temperature dependence \cite{Charlat}. With two S islands as in sample 2, the same 
analysis holds.

\vspace*{0.5 cm}
\epsfxsize=8 cm \epsfbox{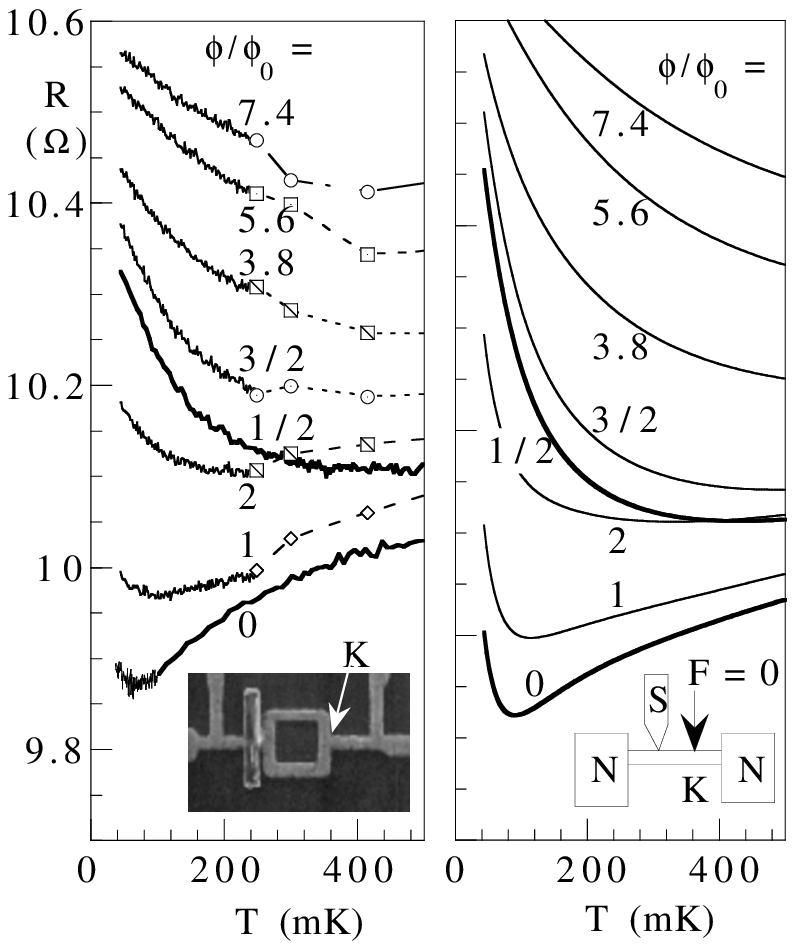}
\begin{center}
\small 
{\bf Fig. 6}, Left : Measured temperature dependence of the resistance of sample 4 at 
different values of the magnetic flux in units of the flux-quantum : $\phi / \phi_0 =$0; 1/2; 
1; 3/2; 2; 3.8; 5.6 ; 7.4. $I_{mes} =$ 200 nA. Right : Calculated 
resistance temperature dependence's for the flux values. The sample model is shown in 
the inset. The arrow shows the location (K) where half a flux-quantum enforces a zero 
pair amplitude. The physical parameters (D, L...) are the measured ones except for the 
width of the N wire $w =$ 50 nm. The zero--magnetic-field phase-breaking length 
$L_\varphi$ is taken to be infinite.
\normalsize
\end{center}

As an additional effect of the magnetic field $H$, the 
phase-memory length $L_{\varphi}$ is renormalized due to the finite width $w$ of the 
Cu wire \cite{Pannetier-Rammal} :
\begin {equation}
L_{\varphi}^{-2}(H)=L_{\varphi}^{-2}(0)+\frac{\pi^{2}}{3} 
\frac{H^{2}w^{2}}{\Phi_{0}^{2}}
\end {equation}
When smaller than the sample length L, the phase-memory length $L_{\varphi}(H)$ 
plays the role of an effective length for the sample. As a result, the resistance minima 
is shifted to higher temperatures when the magnetic field is increased, see Fig. 6.

In the right part of Fig. 6, we show the calculated 
resistance using of Eq. 1-3 in the modelized geometry of Fig. 6 inset in the 
case of a fully transparent interface. The only free parameter is the width of the wires 
which has been adjusted so that the high-field damping of the amplitude of the 
magnetoresistance oscillations is well described by the calculation. The discrepancy 
between the fitted value $w=$ 65 nm and the measured value is attributed to sample 
non-ideality. Our calculation accounts for both the global shape and amplitude of the 
curves and for their behaviour as a function of the magnetic flux. This is particularly 
remarkable in respect with the strong assumptions of the model. One should note that 
qualitative shape and amplitude of the curves are conserved if non-linearized Usadel 
equations or slightly different geometrical parameters are used.

\section{Conclusion}
In conclusion, we have clearly identified the different components of the proximity 
effect in a mesoscopic metal near a superconducting interface. We demonstrated the 
cross-over between the low-temperature Josephson coupling and a energy-sensitive 
conductance enhancement at high temperature. We proved the persistence of the 
proximity effect over the phase-coherence length $L_\varphi$ by observing the 
power-law ($1/T$) temperature dependence of the magneto-resistance oscillations in a 
loop. This effect is much larger than the weak localization correction. We 
observed for the first time the re-entrance of the metallic resistance when all 
energies involved are below the Thouless energy of the 
sample, as predicted in recent theories. Our experimental results are well described by the 
linearized Usadel equations from the quasiclassical theory. We notice that in addition to 
the study of electron-electron interaction in metals \cite{Stoof}, these results are very 
promising for studying the actual efficiency of electron reservoirs \cite{Charlat}.

We thank B. Spivak, F. Wilhem, A. Zaikin and F. Zhou for stimulating 
discussions. The financial support of the Direction de la Recherche et des Technologies 
and of the R\'egion Rh\^one-Alpes is gratefully acknowledged.

\end{document}